\documentclass[twocolumn,showpacs,showkeys,preprintnumbers,amsmath,amssymb,superscriptaddress]{revtex4}
\usepackage{graphicx}% Include figure files
\usepackage{dcolumn}% Align table columns on decimal point
\usepackage{bm}% bold math
\usepackage{xcolor}

\begin{document}

\title{Dual wavefunctions  in two--dimensional quantum  mechanics}

\author{Sergio A. Hojman}
\email{sergio.hojman@uai.cl}
\affiliation{Departamento de Ciencias, Facultad de Artes Liberales,
Universidad Adolfo Ib\'a\~nez, Santiago 7491169, Chile.}
\affiliation{Centro de Investigaci\'on en Matem\'aticas, A.C., Unidad M\'erida, Yuc. 97302, M\'exico}
\affiliation{Departamento de F\'{\i}sica, Facultad de Ciencias, Universidad de Chile,
Santiago 7800003, Chile.}
\affiliation{Centro de Recursos Educativos Avanzados,
CREA, Santiago 7500018, Chile.}

\author{Felipe A. Asenjo}
\email{felipe.asenjo@uai.cl}
\affiliation{Facultad de Ingenier\'ia y Ciencias,
Universidad Adolfo Ib\'a\~nez, Santiago 7491169, Chile.}

\begin{abstract}
It is shown that the Schr\"odinger  equation for a large family of pairs of two--dimensional quantum potentials possess wavefuctions for which the amplitude and the phase are interchangeable, producing two different solutions which are dual to each other. This  is a property 
of solutions with vanishing B\"ohm potential. These solutions can be extended to  three--dimensional systems. We explicitly calculate  dual solutions for physical systems, such as the repulsive harmonic oscillator and the two--dimensional hydrogen atom. These dual wavefunctions are also solutions of an analogue optical system in the eikonal limit. In this case, the potential is related to the refractive index, allowing the study of this two--dimensional dual wavefunction solutions with an optical (analogue) system.
\end{abstract}

%\pacs{04.60 Bc, 98.80 Qc, 11.30 Cp}

\maketitle
\section{Introduction}

Different aspects of quantum mechanical solutions are always worth to be considered, because they may bring new insights on old problems. 
In this manuscript, we study a  set of wavefunctions for which their amplitudes and phases can be interchanged in order
to solve the same Schr\"odinger equation for a family of specific potentials.
It will be shown that this happens in two--dimensional systems when the B\"ohm potential vanishes.

The B\"ohm potential is defined by
\begin{equation}
-\frac{\hbar^2}{2m}\frac{ {\nabla}^2 A }{A}\, , \label{VB}    
\end{equation}
where $A=\sqrt{\psi^*\psi}$, is the amplitude of the wavefunction $\psi$. This B\"ohm potential plays an important role in the quantum theory described by Schr\"odinger equation 
\begin{equation}
\left[-\frac{{\hbar}^2}{2m}\nabla^2 + V - i \hbar \frac{\partial}{\partial t}  \right] \psi = 0\, , \end{equation}
with the potential $V$.
If the wavefunction is written as ${{\psi}}= A\,  \exp({i S/\hbar})$, with a phase $S$, then the Schr\"odinger equation may be written as two coupled equations \cite{book,book2}
\begin{eqnarray}
\frac{1}{2m}\vec {\nabla} S \cdot \vec {\nabla} S - \frac{\hbar^2}{2m}\frac{ {\nabla}^2 A }{A} + V +\frac{\partial S}{\partial t}&=&0\, , \label{HJB1} \\  \frac{1}{m} \vec{\nabla} \cdot (A^2 \vec{\nabla} S)  +\frac{\partial A^2}{\partial t}&=&0\, . \label{cont}    
\end{eqnarray}
These equations are called the Madelung--B\"ohm hydrodynamical version of quantum mechanics.
The B\"ohm potential appears explicitly in Eq.~\eqref{HJB1}. We can also see that equation \eqref{cont} is the continuity (probability conservation) equation.
The importance of describing quantum mechanics in this form is on the recognition of the first equation \eqref{HJB1} as a modified version of the classical Hamilton--Jacobi equation for the potential $V$.  The classical Hamilton--Jacobi equation is modified by the addition of the B\"ohm potential, which  is the only term where Planck's constant $\hbar$ appears in this version of the Schr\"odinger equation. The theory described by Eqs.~\eqref{HJB1} and \eqref{cont}, 
is also called Quantum Hamilton--Jacobi theory,
and has been extensively studied in the context of their differences with respect to Hamilton--Jacobi
equations \cite{book2,Gondran,chin, chin2,bouda,bouda2,ozum,marmo,castro}.

However, in this work we focus in a different feature of quantum mechanics in two--dimensions. We prove that when B\"ohm potential vanishes, there exists a family of potentials for which the wavefunction solutions present a duality of interchangeability between their amplitudes  
and phases.
Below we find the conditions for potential where this
feature be explicitly displayed in the solutions. Furthermore, we show that those two--dimensional solutions correspond to an exact optical analogue (in the geometrical optics limit), for which the potential is closely related to a refractive index. Therefore, these particular solutions can be studied optically. We present some optical systems which are suitable for this purpose.

\section{Duality solutions in two dimensions}

Let us consider
 a special family of two--dimensional quantum potentials $V_1 (x,y)$, such that if one wave--function ${{\psi}_u}$, given by
\begin{equation}
{{\psi}_u} \equiv A\ \exp\left(\frac{i}{\hbar}S\right)\, , \label{psiA}     \end{equation}
(with $A, S \in \mathbb{R}$) solves the Schr\"odinger equation
\begin{equation}
\left[-\frac{{\hbar}^2}{2m}\nabla^2 + V_1 (x,y) - i \hbar \frac{\partial}{\partial t}  \right] \psi_u = 0\, ,\label{schr1}
\end{equation}
then, a different wave--function ${{\psi}_v}$, given by 
\begin{equation}
{{\psi}_v} \equiv S\ \exp\left(\frac{i}{\hbar}A\right)\, , \label{psiS}       \end{equation}
solves the same Schr\"odinger equation \eqref{schr1}. For that quantum potential, the duality of the wavefunction solutions of Eq.~\eqref{schr1} correspond to the interchangeability between the amplitude and the phase in wavefunctions \eqref{psiA} and \eqref{psiS}.

The family of potentials which allows this kind of wavefunction solutions
is composed of couples of related potentials $V_1 (x,y)$ and $V_2 (x,y)$ \cite{hcnr}, such that the general solutions to their respective Schr\"odinger equation \eqref{schr1} may be written in terms of the solutions to the other Schr\"odinger equation
\begin{equation}
\left[-\frac{{\hbar}^2}{2m}\nabla^2 + V_2 (x,y) - i \hbar \frac{\partial}{\partial t}  \right] \psi_{u,v} = 0, \label{schr2}
\end{equation}
where $V_1 (x,y)$ [$V_2 (x,y)$] is completely determined by $V_2 (x,y)$ [$V_1 (x,y)$].
The family of potentials $V(x,y)$ that fulfill this condition is
defined by \cite{hcnr}
\begin{equation}\label{factorV}
{\nabla}^2 \log V(x,y)=0\, .
\end{equation}

In the following sections we prove the above statements, first for time--independent two dimensional quantum mechanics, to later extend this result for time-dependent solutions in three dimensions.

\subsection{Time--independent two--dimensional system}

Consider a solution to a two--dimensional Schr\"odinger equation for a special potential $V(x,y)$ written in terms of a two--dimensional time--independent wavefunction
\begin{equation}
{{\psi}_u} (x,y) \equiv A(x,y)\ e^{i S(x,y)/\hbar}, \label{psia}     
\end{equation}
where $A(x,y)$ and $S(x,y)$ are real functions of two variables $x$ and $y$.
The real and imaginary parts of the Schr\"odinger equation satisfy the two--dimensional version of Eqs.~\eqref{HJB1} and \eqref{cont}.

%%%%%%%%%%

Now, consider the two--dimensional functions $u(x,y)$ and $v(x,y)$ as the real and imaginary parts of a holomorphic function $f(z)$ respectively, 
\begin{equation}\label{functionfgeneraluv}
  f(z)= u(x,y)+iv(x,y)\, ,  
\end{equation} 
 written in terms of a complex variable $z$
 \begin{equation}\label{complexvariablez}
      z=x+iy= r \exp(i\theta)\, ,
 \end{equation}
 where
$r=\sqrt{x^2+y^2}$, and $\theta=\arctan\left({y}/{x}\right)$.
 As $f$ is holomorphic, the functions $u(x,y)$ and $v(x,y)$ satisfy the Cauchy--Riemann conditions
 \begin{eqnarray}
 \frac{\partial u}{\partial x}&=&\frac{\partial v}{\partial y}\, ,\nonumber\\
 \frac{\partial u}{\partial y}&=&-\frac{\partial v}{\partial x}\, .
 \end{eqnarray}
 It is well known (and straightforward to prove using the Cauchy--Riemann conditions) that these two--dimensional functions $u(x,y)$ and $v(x,y)$ fulfil
 \begin{eqnarray}
 {\nabla}^2 u&=&0={\nabla}^2 v\, ,\label{lapl}\\
0&=&{\vec{\nabla}} u \cdot {\vec{\nabla}} v\, , \label{orthog}\\
{\vec{\nabla}} u \cdot {\vec{\nabla}} u &=& {\vec{\nabla}} v \cdot {\vec{\nabla}} v\, . \label{potential}
\end{eqnarray}
Consider now a (factorizable) family of potentials $V(x,y)$ defined by Eq.~\eqref{factorV}.
Any member of such family may be factorized as 
\begin{equation}\label{ecuaforS}
-2mV(x,y) =g(z){\bar g}({\bar z})\, .
\end{equation}
Therefore, Eqs.~\eqref{HJB1} and \eqref{cont} written for any member of the factorizable family in two--dimensions, are solved by the wavefunction \eqref{psia} with $A(x,y) = u(x,y)$ and $S(x,y) = v(x,y)$, such that the holomorphic function $g(z)$ is defined by
\begin{equation}
g(z)=\frac{df(z)}{dz}\,  \label{defg}
\end{equation}
because of \eqref{potential}.

Note that as a consequence Eqs. \eqref{lapl}, and \eqref{orthog}, the B\"ohm potential \eqref{VB} vanishes and the continuity equation \eqref{cont} is identically satisfied. Furthermore, of course, the wavefunction \eqref{psia} with $A(x,y) = v(x,y)$ and $S(x,y) = u(x,y)$ (which is dual to the previously constructed solution, with amplitude and phase interchanged) also satisfies Eqs.~\eqref{HJB1} and \eqref{cont} for the same potential because of Eq.~\eqref{potential}.

The time--independent dual solutions exhibited above, really apply to two--dimensional optics because the Hamilton--Jacobi equation for zero energy (recall that $\partial S/ \partial t = 0$), is equivalent to the optical eikonal equation with a refractive index $n(x,y)$ given by
\begin{equation}\label{refractive}
n^2 (x,y) = - 2 m V(x,y)= \  g(z){\bar g}({\bar z}).
\end{equation}
As a consequence, dual solutions are only found for negative potentials given by \eqref{ecuaforS}.

\subsection{A property of holomorphic functions}

The above two--dimensional solution can be shown to emerge as a property of general holomorphic functions. Thus, this solution is a geometrical consequence of the two--dimensions in which the system is restricted to evolve. In terms of the general holomorphic function 
\eqref{functionfgeneraluv}, the above proposed wavefunction solutions are written as
\begin{eqnarray}
\psi_u &=& \frac{f+\bar{f}}{2 } \exp\left(\frac{f-\bar{f}}{2 \hbar}\right)\,  , \label{solA}\nonumber\\
\psi_v &=& \frac{f-\bar{f}}{2 i } \exp\left(i \frac{f+\bar{f}}{2 \hbar}\right)\, , \label{solS}
\end{eqnarray}
as $u=(f+{\bar f})/2$ and $v=(f-{\bar f})/2i$.
Then, we readily get
\begin{eqnarray}\label{eq21geometricholo}
\nabla^2\psi_{u,v}&=& 4\, \frac{\partial^2 \psi_{u,v}}{\partial z \partial \bar {z}}=4\, \frac{\partial^2 \psi_{u,v}}{\partial f \partial \bar {f}} \, g\,  {\bar{g}}\nonumber\\
&=&- \frac{1}{\hbar^2 }g\,  {\bar{g}}\,  \psi_{u,v}\, .
\end{eqnarray}
with $g$ defined in  \eqref{defg}. We have used the well--known result for a two--dimensional space
$\nabla^2 \equiv {\partial_x^2} +{\partial_y^2} = 4 {\partial^2_{z,  \bar {z}}}$,
in terms of complex variable \eqref{complexvariablez}.

Eq.~\eqref{eq21geometricholo} is equivalent to the Schr\"odinger equations Eqs. \eqref{schr1} and \eqref{schr2} for zero energy and potential \eqref{ecuaforS}.
Therefore our result may also be understood as a general property of some functions ($\psi_u$ and $\psi_v$) constructed from an arbitrary holomorphic function $f(z)$ and its complex conjugate $\bar{f}(\bar{z})$. Note that the Bohm potential plays no role in this  approach. This gives a geometrical explanation of the vanishing of  Bohm potential when these dual functions are considered.

\subsection{Time--dependent three dimensional extension}

We can generalize the previous case to time--dependent systems in three--dimensions.
Consider a wavefunction  that propagates in a ${\zeta}$--direction (perpendicular to the previous ones), with energy $E$. It has the form
\begin{equation}\label{wavefunctiondualu}
\psi_{u}(x,y,\zeta,t)= u(x,y) \exp\left(\frac{i}{\hbar}\left[ v(x,y)+k_\zeta\, \zeta-Et\right]\right)\, ,
\end{equation}
where $k_\zeta$ is the wavenumber associated to the ${\zeta}$--direction. Wavefunction \eqref{wavefunctiondualu} solves Eqs.~\eqref{HJB1} and \eqref{cont} for the same two--dimensional potential $V(x,y)$, provided the energy is given by
\begin{equation}
E=\frac{k_\zeta^2}{2m}\, .
\end{equation}
Due to the  Cauchy--Riemann conditions, the same functions $f(z)$ and $g(z)$ allow us to show that the dual wavefunction
\begin{equation}\label{wavefunctiondualv}
\psi_v (x,y,\zeta,t) = v(x,y) \exp\left(\frac{i}{\hbar}\left[u(x,y)+k_\zeta\, \zeta-Et \right]\right)\, .
\end{equation} 
also solves Eqs.\eqref{HJB1} and \eqref{cont}.

A more complete solution can be found if the energy  $E$ has the form
$E={k_\zeta^2}/({2m})+{\cal E}$,
with a constant energy shift ${\cal E}$, while the potential fulfill now
\begin{equation}\label{potencialyepslion}
    -2m\left[V(x,y)-{\cal E}\right] =g(z){\bar g}({\bar z})\, .
    \end{equation}

\section{Solutions for dual wavefunctions}

Based in the above two--dimensional system, with the dual wavefunction solutions, in this section we explore different examples of potentials  whose Schr\"odinger  equations may be solved exactly. We start from simple algebraic choices for the holomorphic function $g$, to finally solve specific physical models, also showing how they correspond to optical systems with different refractive indices in the following section.

\subsection{A general solution}

Let us consider the general two--dimensional potential 
given by
\begin{equation}\label{potentialgeneraly}
V(x,y)=-\alpha\, r^n\, ,
\end{equation}
for $\alpha, n \in \mathbb{R}$, with $\alpha>0$. 
For this potential, the solution of Eq.~\eqref{ecuaforS} is given by
\begin{equation}\label{generalgtot}
    g(z)=\sqrt{2 m\alpha}\,  z^{n/2}\, ,
\end{equation}
and by Eq.~\eqref{defg} we have
\begin{equation}\label{generalftot}
    f(z)=\frac{2\sqrt{2m\alpha}}{n+2}z^{1+n/2}\, .
\end{equation}

\begin{widetext}
Therefore, the dual solutions are given in general by
\begin{eqnarray}\label{generalwavefunctionN}
\psi_u (x,y) &=& \frac{2\sqrt{2m\alpha}}{n+2}r^{1+n/2}\cos\left[\frac{(n+2)\theta}{2}\right] \exp\left(\frac{i}{\hbar} \frac{2\sqrt{2m\alpha}}{n+2}r^{1+n/2}\sin\left[\frac{(n+2)\theta}{2}\right] \right)\, ,\nonumber\\
\psi_v (x,y) &=& \frac{2\sqrt{2m\alpha}}{n+2}r^{1+n/2}\sin\left[\frac{(n+2)\theta}{2}\right] \exp\left(\frac{i}{\hbar} \frac{2\sqrt{2m\alpha}}{n+2}r^{1+n/2}\cos\left[\frac{(n+2)\theta}{2}\right] \right)\, .
\end{eqnarray} 
\end{widetext}
Clearly, these solutions are valid for $n\neq -2$. The case  $n=2$ is discussed below independently. 
Interestingly,
it is straightforward to prove that wavefunctions \eqref{generalwavefunctionN} are single-valued for 
\begin{eqnarray}\label{singlevalueangle}
\theta\rightarrow \theta+\frac{4\pi}{n+2}\, .
\end{eqnarray}

In order to evaluate explicitly the above general solution, in the following sections we focus in particular cases for different $n$ values, showing that these dual solutions can be relevant in standard quantum mechanical scenarios in two dimensions.

\subsection{The simplest holomorphic function}
\label{simplestsolutionholo}

Consider $n=0$ in  \eqref{potentialgeneraly} for a constant potential. 
In this case, $f=\sqrt{2m\alpha}\, z$, and 
\begin{eqnarray}\label{waef1a}
\psi_u(x,y) &=& \sqrt{2m\alpha}\, x\,  \exp\left(\frac{i}{\hbar}\sqrt{2m\alpha} y\right)\, ,\nonumber\\
\psi_v(x,y) &=&\sqrt{2m\alpha}\, y\,  \exp\left( - \frac{i}{\hbar}\sqrt{2m\alpha} x\right)\, .
\end{eqnarray}
 The first solution corresponds to a function $f=\sqrt{2m\alpha}z$, whereas the second one is for $f=-i\sqrt{2m\alpha}z$.
 It is straightforward to show by inspection that the dual wavefunctions \eqref{waef1a} satisfy 
Schr\"odinger equation \eqref{schr1}.
These solutions are single--valued for $\theta\rightarrow\theta+2\pi$.

\subsection{Repulsive harmonic oscillator in two dimensions}

Consider $n=2$ in potential \eqref{potentialgeneraly}, to get the repulsive harmonic oscillator
\begin{equation}\label{potencrepulsivo}
    V=-\alpha\,  r^2\, .
\end{equation}
Thus, we find that 
\begin{equation}\label{solurepuslive}
    g(z)=\sqrt{2m\alpha}\,  z\, ,
\end{equation}
and similarly we find from \eqref{generalftot} that
\begin{equation}
    f(z)=\sqrt{\frac{m\alpha}{2}}\, z^2\, .
\end{equation}
Thereby, the dual wavefunctions
\begin{eqnarray}\label{waef2b}
\psi_u(x,y) &=&\sqrt{\frac{m\alpha}{2}}r^2\cos(2\theta)\exp\left[\frac{i}{\hbar}\sqrt{\frac{m\alpha}{2}}\, r^2\sin(2\theta) \right]\, ,\nonumber\\
\psi_v(x,y) &=&\sqrt{\frac{m \alpha}{2}}\, r^2\sin(2\theta)\exp\left[\frac{i}{\hbar} \sqrt{\frac{m\alpha}{2}}r^2\cos(2\theta)\right]\, ,
\end{eqnarray}
are solutions of Schr\"odinger equation \eqref{schr1} for  potential \eqref{potencrepulsivo}.
These solutions are single--valued for $\theta\rightarrow\theta+\pi$.

\subsection{Hydrogen atom in two dimensions}
\label{hydrogenatomsection}

Let us consider $n=-1$ in \eqref{potentialgeneraly}. Thus we find
the  potential for a two--dimensional hydrogen atom
\begin{equation}\label{hydrogenatompotential}
    V=-\frac{\alpha}{r}\, .
\end{equation}
 By using \eqref{generalgtot} we find 
\begin{equation}\label{ghydrogat}
    g(z)=\sqrt{\frac{2m\alpha}{z}}\, ,
\end{equation} and thus 
\begin{equation}
    f(z)=2\sqrt{2m\alpha z}\, .
\end{equation} 
This allows us to find the solutions for the dual wavefunctions
\begin{eqnarray}\label{waef3a}
\psi_u(x,y) &=&2\sqrt{2m\alpha\, r} \cos\left(\frac{\theta}{2}\right) \exp\left[\frac{2i\sqrt{2m\alpha r}}{\hbar}\sin\left(\frac{\theta}{2}\right)\right]\, ,\nonumber\\
\psi_v(x,y) &=&2\sqrt{2m\alpha\,  r} \sin\left(\frac{\theta}{2}\right) \exp\left[\frac{2i\sqrt{2m\alpha r}}{\hbar}\cos\left(\frac{\theta}{2}\right)\right]\, .\nonumber\\
\end{eqnarray}

Anew, the above wavefunctions are single--valued for $\theta\rightarrow\theta+4\pi$.

\subsection{The exceptional potential for $n=-2$}

For this case
\begin{equation}\label{anom}
    V(x,y)= -  \frac{{\alpha}}{r^2}\, .
\end{equation}
This potential is remarkable in its own right because it constitutes one of the few known examples of anomalies in quantum mechanics \cite{coon}. It is a simple matter to get
\begin{eqnarray}\label{r-2}
    g(z)&=& \frac{\sqrt{2m\alpha}}{z}\, ,\nonumber\\
    f(z)&=&  \sqrt{2m\alpha}\ \log(z)\, .
\end{eqnarray}
And then, both dual wavefunctions are 
\begin{eqnarray}\label{wf1}
    \psi_u (x,y)&=& \log r\ \exp(i \theta)\, ,\nonumber\\
    \psi_v(x,y)&=&\theta\ \exp(i\log r)= \theta\ r^i\, .
\end{eqnarray}
Notice here that solutions \eqref{generalwavefunctionN} do not apply. In fact, wavefunction $\psi_u$ is single--valued for $\theta\rightarrow\theta+2\pi$, whereas $\psi_v$ does not have such property.

\section{Analogy with refractive indices}

For any $n\in \mathbb{R}$, solutions of potential \eqref{potentialgeneraly} can be found. However, it is important to remark another feature of the dual solutions found here. The two-dimensional system can be studied as an analogue model for lenses with specific refractive indices.

If we choose the refractive index $n(x,y)$ to be real, then $n^2$ is positive [recall Eq.~\eqref{refractive}]
which, in turn, means that a two--dimensional dual quantum mechanics system can be made analogous to an optical model for negative potentials only, $V<0$.

In the following sections, we present examples where refractive indices can be used to model dual quantum mechanical systems.

\subsection{Eaton lens and hydrogen atom in two dimensions}

An Eaton, lens with refraction angle of $\pi$, can be described by the refractive index \cite{kim,zeng,martin}
\begin{equation}
    n(x,y)=\sqrt{\frac{2a}{r}-1}\, ,
\end{equation}
where $a$ is the radius of the lens, and $r\leq a$.

Using Eq.~\eqref{potencialyepslion}, we find the associated potential 
\begin{equation}
    V(x,y)=-\frac{a}{m r}\, ,
\end{equation}
with an energy shift  ${\cal E}=-1/2m$.
This potential corresponds to the choices $\alpha=a/m$, and $n=-1$ in \eqref{potentialgeneraly}. Therefore, we are able to find  the solution \eqref{ghydrogat} for a hydrogen atom in two dimensions
\begin{equation}
    g(z)=\sqrt{\frac{2a}{z}}\, .
\end{equation}
In this form, an Eaton lens, with a refraction angle of $\pi$, is equivalent to the dual quantum mechanical system of a hydrogen atom in two dimensions with energy $E=(k^2-1)/2m$.

\subsection{More general Eaton Lens}

For any refraction angle $\phi$, an Eaton lens can be approximated as \cite{kim}
\begin{equation}
    n(x,y)\approx \left(\frac{2a}{r}-1\right)^{\phi/(\phi+\pi)}\, .
\end{equation}
Close to the center of the spherical lens ($r\ll a$), we have an approximate refractive index
\begin{equation}
    n(x,y)\approx \left(\frac{2a}{r}\right)^{\phi/(\phi+\pi)}\, ,
\end{equation}
which allows us to find the potential
\begin{equation}
    V(x,y)=-\frac{1}{2m}\left(\frac{2a}{r}\right)^{2\phi/(\phi+\pi)}\, ,
\end{equation}
with ${\cal E}=0$.
This corresponds to a general potential \eqref{potentialgeneraly} with $\alpha=(2a)^{\phi/(\phi+\pi)}/2m$, and
$n=-2\phi/(\phi+\pi)$.
With Eqs.~\eqref{generalgtot} and \eqref{generalftot} we find
\begin{eqnarray}
g(z)&=&(2 a)^{\phi/(\phi+\pi)}z^{-\phi/(\phi+\pi)}\, ,\nonumber\\
f(z)&=&\frac{(\phi+\pi)}{\phi}(2 a)^{\phi/(\phi+\pi)}z^{\pi/(\phi+\pi)}\, ,
\end{eqnarray}
from where the dual wavefunctions can be constructed.

\subsection{General monomial refractive index}

By using the general potential \eqref{potentialgeneraly}, we can construct the general refractive index for an optical system be analogous to  two-dimensional dual quantum mechanics.
This is
\begin{equation}
    n(x,y)= \sqrt{2m\alpha\, r^n}\, .
\end{equation}

This result implies that all the  previous  mentioned refractive indices can be studied optically, whereas some other ones, such as L\"uneburg or Maxwell lenses \cite{martin}, are excluded from this treatment of optical analogies.

\section{Conclusions}

We have presented a large family of quantum mechanical (and their related optical) problems which possess a remarkable duality property, in the sense that in some of their solutions amplitudes and phases are interchangeable. This implies that this kind of quantum mechanical systems affords solutions in which particles with the same energy move with different momenta. This can be proved by two alternative approaches, one based on the condition of vanishing Bohm potential, while the other is based on two--dimensional properties of arbitrary holomorphic functions.

These solutions with the duality properties, include interesting physical systems. An example is the two--dimensional hydrogen atom solution described in Sec.~\ref{hydrogenatomsection}. Solutions of the two--dimensional hydrogen atom potential
\eqref{hydrogenatompotential} are known \cite{yangguo,Parfitt}, although they have a non--vanishing B\"ohm potential, being different to ours.

On the other hand, the fact that quantum mechanical and optical problems may be treated in a unified way allows for the possibility of studying electron propagation in some potentials using analogous experiments with light propagation, which sometimes are easier to handle. This can be achieved in physical optical systems, as using  Eaton lenses, for instance. This opens new types of quantum--mechanical experiments at classical level.

Lastly, the relation among
 wavefunctions associated to two--dimensional potential \eqref{potentialgeneraly}, transformation \eqref{singlevalueangle}, and anyon statistics \cite{lein,wilc,lardo} is currently under research.

%%%%%%%%%%%%%%%%%%%%%%%%%%%%%

\end{document}